\documentclass[11pt,a4paper]{article}
\usepackage{amsmath,amssymb,amsfonts,textcomp}
\usepackage{verbatim}
\usepackage{lscape}

  \usepackage[
  pdftex,
  hyperindex,
  pagebackref,
  colorlinks,
  linkcolor=blue,
  menucolor=blue,
  pagecolor=blue,
  urlcolor=blue,
  citecolor=blue,
  pdftitle={Neuroelectrics Barcelona SL -TN},
  pdfauthor={NE},
  pdfcreator={LaTeX}
  ]{hyperref}

  \usepackage{thumbpdf}
  \usepackage{url}
  \usepackage[pdftex]{graphicx}
  \DeclareGraphicsExtensions{.pdf,.jpg,.mps,.png}

\usepackage[margin=10pt,font=small,labelfont=bf, labelsep=endash]{caption}

\usepackage{fancyhdr}
\usepackage{multicol}

\def\_#1 {_{\rm #1}}

\catcode`\@=11

\@addtoreset{equation}{section}
\def\theequation{\arabic{section}.\arabic{equation}}
\catcode`\@=11

\def\section{\@startsection{section}{1}{\z@}{3.5ex plus 1ex minus
   .2ex}{2.3ex plus .2ex}{\large\bf}}

\def\eqnarray{\let\@currentlabel=\theequation\refstepcounter{equation}
    \global\@eqnswtrue
    \global\@eqcnt\z@\tabskip\@centering\let\\=\@eqncr
    $$\halign to \displaywidth\bgroup\@eqnsel\hskip\@centering
      $\displaystyle\tabskip\z@{##}$&\global\@eqcnt\@ne
       \hfil${{}##{}}$\hfil
      &\global\@eqcnt\tw@ $\displaystyle\tabskip\z@{##}$\hfil
       \tabskip\@centering&\llap{##}\tabskip\z@\cr}
\def\lefteqn#1{\hbox to 4\arraycolsep{$\displaystyle #1$\hss}}
\begin{document}

%%%%%%%%%%%%% Start title page %%%%%%%%%%%%%

\pagestyle{myheadings}
\markboth{NE Technical Note TN0008}{Neuroelectrics Barcelona SL - TN0008}

\noindent
{\large\bf Application of the reciprocity theorem to EEG inversion and optimization of EEG-driven tCS (tDCS, tACS and tRNS)}

\vspace{0.5cm} \noindent
{\normalsize {\bf Author:} Giulio Ruffini (giulio.ruffini@neuroelectrics.com) }\\ \noindent
{\normalsize {\bf Date:} June 10 2015}\\ \noindent
%{\normalsize {\bf Project:} Starstim }\hfill {\large {\bf Status:} Open}\\

\vspace{0.3cm} \noindent
\line(1,0){400}

%%%%%%%%%%%%% End title page %%%%%%%%%%%%%

\section*{Summary}
       Multichannel transcranial current stimulation systems offer the possibility of EEG-guided optimized brain stimulation. 
       In this brief technical note I  explain how it is possible to use transcranial current stimulation tCS (which includes tDCS, tACS and tRNS among others\footnote{ Ruffini G, Wendling F, Merlet I, Molaee-Ardekani B, Mekonnen A, Salvador R, Soria-Frisch A, Grau C, Dunne S, Miranda PC, Transcranial current brain stimulation (tCS): models and technologies, IEEE Trans Neural Syst Rehabil Eng. 2013 May;21(3):333-45. }) electric field realistic brain models to create a forward ``lead-field'' matrix and, from that,  an EEG inverter for cortical mapping.  Starting from EEG we show how to generate 2D cortical surface dipole fields that could generate the observed EEG electrode voltages.  The main tool is the reciprocity theorem derived by Helmholtz. The application of reciprocity for the generation of a forward mapping matrix (lead field matrix as is sometimes known) is well known [Rush1969\footnote{Rush and Driscoll, 1969, IEEE Trans. Biom. Eng.}], but here we will use it in combination with the realistic head models of [Miranda2013\footnote{Miranda et al., NeuroImage 70 (2013) 48Ð58}] to provide cortical mapping solutions compatible with realistic head model tCS optimization.
       
       I also provide a generalization of the reciprocity theorem [Helmholtz1853\footnote{H. Helmholtz, "Uber einige Gesetz der Vertheilung elektrischer Strdme in korperlichen Leitern, mit Anwendung auf die thierisch- elektrischen Versuche," Ann. Phys. Chem., ser. 3, vol. 29, pp. 211-233 and 353-377, 1853.}] to the case of multiple electrode contact points and EEG dipole sources, and discuss its uses in non-invasive brain stimulation based on EEG. This can be used to guide for optimization of transcranial current stimulation with multiple channels based on EEG.
This, as far as I know, is a novel result.
 
 \clearpage
\tableofcontents
\clearpage
%%%%%%%%%%%%%%%%%%

%%%%%%%%%%%%%%%%%%%%%%
\clearpage 
\section{Introduction}
The motivation of this note is to develop ways to derive targeted multichannel transcranial current stimulation (tCS) protocols from EEG data.  Since both  EEG and tCS are mostly cortically relevant, we would like to develop ways to compare the generated electric fields of tCS with EEG dipole cortical field sources.

There are now many groups developing realistic, FEM-based models of current propagation in the human head, solving Poisson's equation. In the Starstim software\footnote{ http://www.neuroelectrics.com} we have now integrated such realistic head modeling of the generated electric fields by non-invasive multichannel brain stimulation, based on the work of [Miranda2013]. 

How can we use such current propagation (electric field) models in to infer an EEG dipole forward model?  They are certainly related by the reciprocity theorem and by the (not-unrelated) fact that are both governed by Poisson's equation. Once we derive a realistic forward model we can develop inverse modeling approaches, to go from scalp to source space. 

In this note I  concentrate first on the problem of cortical mapping, i.e., finding a dipole distribution on the cortical surface that will generate the EEG pattern. Although the method described can be used to create volume dipole vector field forward operators, given the fact that in our own current multichannel systems the number of electrodes is limited to 32, and the issues with volume inverse modeling (a very ill-posed problem) it is reasonable to work with more constrained models: we fix the locations of the dipoles to the cortical surface and their orientation normal to it, reflecting our current understanding that EEG sources are mostly cortical and normally aligned to the surface (pyramidal neuron populations). Surface mesh points in our models on the cortical surface are of the order of 35,000.  

The transformation from voltage space to dipole space is linear. The main task is to find the linear transformation (the ``lead field'' matrix).
What follows is based on the reciprocity theorem. Once we derive a cortically mapper, tCS montages can be optimized to target the desired EEG features in cortical space. 

Finally, I provide also an extension of this theorem to include the case of multiple contact points and current sources. This can be used to guide for optimization of transcranial current stimulation with multiple channels based on EEG.

\section{Reciprocity}

The reciprocity theorem states that there is a relationship between the following variables:
\begin{itemize}
\item $V^{(1)}_{ab}$ and $\vec{J}^{(1)}$: $V^{(1)}_{ab}$ is the observed potential difference between two points in the scalp due to a lone dipole current source in the brain, $\vec{J}^{(1)}$ (in a volume $\delta V$)
\item 
$I^{(2)}_{ab}$ and $\vec{J}^{(2)}$: $I^{(2)}_{ab}$ is an imposed, stimulation, current between these two points in the scalp and the resulting current density in the brain, $\vec{J}^{(2)}$
\end{itemize}

The reciprocity relation is
$$
 \sigma  V^{(1)}_{ab} \, I^{(2)}_{ab} = - \vec{J}^{(1)} \cdot  \vec{J}^{(2)} \delta V
$$
or
$$
 V^{(1)}_{ab} \, I^{(2)}_{ab} = - \vec{J}^{(1)} \cdot  \vec{E}^{(2)} \delta V
$$
 
 Dropping the reference and other indices, we can just write:
 
 \begin{equation}
% V_{a} \, I_{a} = - \vec{J}(x) \cdot  \vec{E}(x) \delta V
 \color{blue}  V_{a} \,  \color{red} I_{a} = - \color{blue} \vec{J}(x) \cdot  \color{red} \vec{E}(x) \color{black}  \delta V
\end{equation}
 (see the figure below). This is nicely explained by [Plonsey1963\footnote{Plonsey R., (1963) Reciprocity applied to volume conductors and the EEG. IEEE Trans. Biomed. Electron. 10:9Ð12}] paper and later book.
 
 With regards to units, if we input uV and we work with 1 mA current as a reference, the output will be in nA m.
 
 %Also, I note that since we have now changed our E field convention in StimPlanner to an inward normal for display purposes, we should probably change the sign to display outgoing dipoles as positive. This is implemented in the Matlab code.
 
The problem we wish to address is: what is the voltage $V_{ab}$ at a point $b$ w.r.t. point $a$ due to a lone dipole $\vec{J}^{(1)}$?  The answer is:
$$
V^{(1)}_{ab} = - {\vec{J}^{(1)} \cdot  \vec{E}^{(2)} \delta V \over  I^{(2)}_{ab}}
$$
For more dipoles the superposition principle applies,
$$
V^{(1)}_{ab} = - \sum_n  {\vec{J_n}^{(1)} \cdot  \vec{E_n}^{(2)} \delta V \over  I^{(2)}_{ab}}
$$
where $n$ refers to voxel location ID.

Now, in general, the volume corresponding to each dipole will not be the same.  A more refined equation is 
$$
V^{(1)}_{ab} = - \sum_n  {{ \vec{E_n}^{(2)}}  \cdot  \vec{J_n}^{(1)}   \over  I^{(2)}_{ab}} \delta V_n .
$$

Now, we have computed the solutions to all combinations of electrodes in our EEG cap with respect to  Cz.  Using this expression we can thus compute the voltage of any channel location s w.r.t Cz. 

We can now define the algorithm for the forward model.
\begin{enumerate}
\item Put dipole where you want (a 3D vector). 
\item  Dot product with the electric field $\vec{E_n^0}^{(2)} $produced by a the reciprocal montage at $a$ and $b$ using the realistic head model with a unitary current (I=1). 
\item Multiply by volume of dipole. 
\end{enumerate}
You now have the potential at site $a$ w.r.t $b$ ($V_{ab} = V_a -V_b$).

$$
V^{(1)}_{ab} = - \sum_n  {{ \vec{E_n^0}^{(2)}}  \cdot  \vec{J_n}^{(1)}   } \delta V_n .
$$

In a triangular mesh, each mesh triangle can be associated to its defining  three mesh points. So  if we define the area of a mesh point to be the area of all the triangles it  touches, and then this area over all mesh points, each triangle will show up three times. Thus,  to account for dipole area/volume, we can properly assign an area to each mesh $m$ point by the formula
$$
A_m= {1\over 3} \sum_{\mbox{\tiny triangles $t$ touched by meshpoint $m$}} A^{T_m}_t
$$

Do this for every single component of the dipole sources you want your model to represent (e.g., all those in the cortex) ... 

Algorithmically, we have as many basis functions as electrodes minus 1, defining currents from Cz to other points. We need to loop around them, and then loop over dipole space (3N) to create the matrix.  The output will be a 26 x 3N matrix, the famous ``lead'' or forward-mapping  $K$ matrix. 

An  option for use here is to focus on the cortical surface and only on normally oriented dipoles. This will make the problem much more amenable. Interesting, only the normal E fields generated by stimulation are relevant according to first order models of neuron electric field interaction. 

\subsection{Inversion with EEG data as constraints}

Once we have the forward mapping  $K$ matrix, we need to smooth the problem for inversion.  An approach I've used in the past is to minimise the ``curvature" of the solution subject to the data constraints.  See [Ruffini2002\footnote{  ``Spherical Harmonics Interpolation, Computation of Laplacians and Gauge Theory" at http://arxiv.org/abs/physics/0206007}].

Here we do the same, with a different  implementation of the Laplacian operator which can be used on a mesh. 

A simpler approach we can employ here is to use the (over the surface) distance matrix across cortical surface points. The matrix to use for regularisation would be
$$
R(x,x')=\delta(x,x')- {d(x,x')^{-q} \over \sum_{x\neq x'}d(x,x')^{-q}} 
$$
The equation $RJ=0$ says the ``the value of J at a point is equal to the average of points nearby weighted by distance''. We can start with $q=2$, 
$$
R(x,x')=\delta(x,x')- {d(x,x')^{-2} \over \sum_{x\neq x'}d(x,x')^{-2}} 
$$

More generally, we can write
$$
R(x,x')=\delta(x,x')- {f(d(x,x')) \over \sum_{x\neq x'}f(d(x,x'))} 
$$
E.g., we can use
$$
f(d(x,x')) = \exp [  -d(x,x')/ d_0 ] 
$$

We can follow the work in [Ruffini2002]. We aim to minimise "curvature" here defined by $R$ subject to the constraints that the EEG data impose ($V=KJ$), which are added using a family of Lagrange multipliers:
$$
\chi={1\over2} J^T  R^T R J +\lambda^T \cdot (V- K J)
$$
and the solution  to this problem is given by
$$
J=\left[  R^TR + K^T K - K^T (K K^T)^{-1} K  R^TR \right] ^{-1} K^T V
$$
as explained in the aforementioned technical note on arxiv.

%%%%%%%%%
\subsection{Inversion with standard spatial regularisation}
A related approach is to start from a functional in which both data and curvature constraints appear treated equally. Mathematically this means $\lambda$ is input by hand and not solved for. The solution is
$$
J=[ K^T K + \alpha R^T R ]^{-1} K^T V
$$
where the parameter $\alpha$ is input by hand.

Another option is Tikhonov regularisation (using $I$ instead of $R$), 
$$
J=[ K^T K + \alpha I]^{-1} K^T V
$$
where the parameter $\alpha$ is input by hand.

Implicit referencing to Cz is used.

\begin{figure}[b!] \hspace{-0cm}  %\vspace{-5 cm}
        \includegraphics[width=14cm,angle=0]{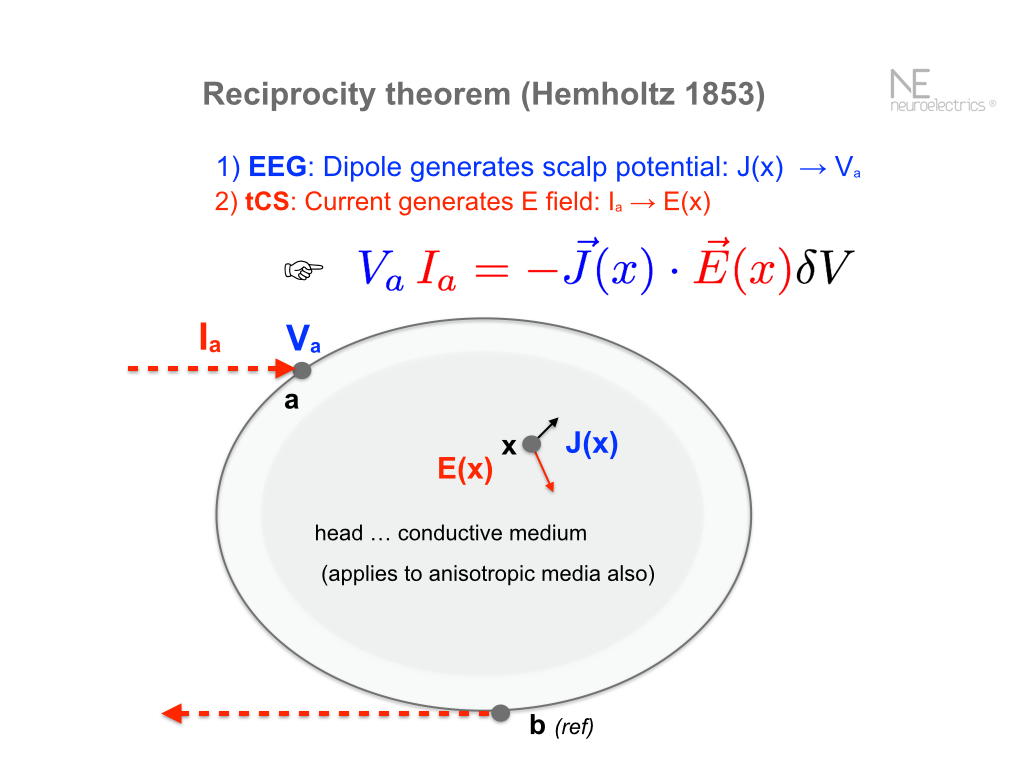}        
    \caption{Top: Reciprocity relation symbols and actors } 
    
  \label{fig:reciprocity}

\end{figure}

\begin{figure}[b!] \hspace{-0cm}  %\vspace{-5 cm}

        \includegraphics[width=12cm,angle=0]{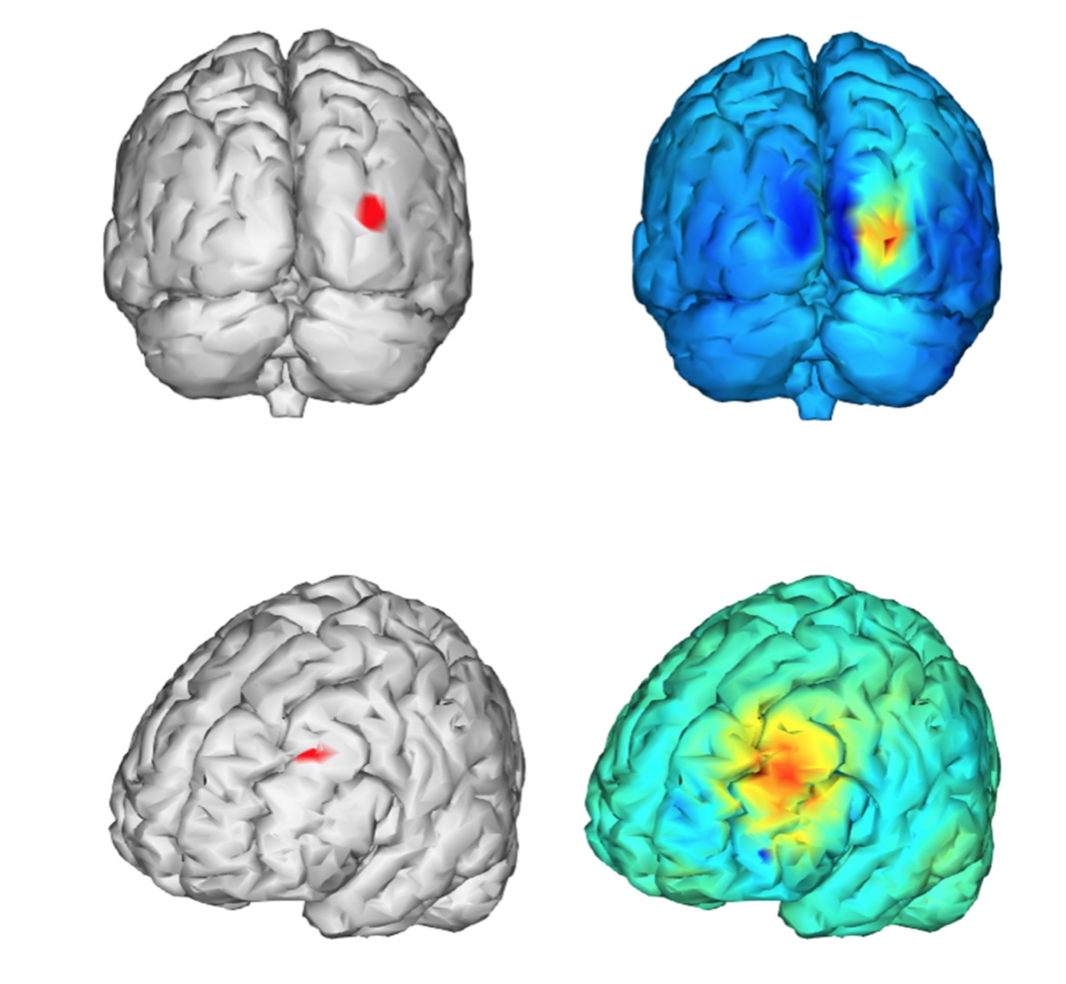}
        
              \hspace{2 cm}   \includegraphics[width=8cm,angle=0]{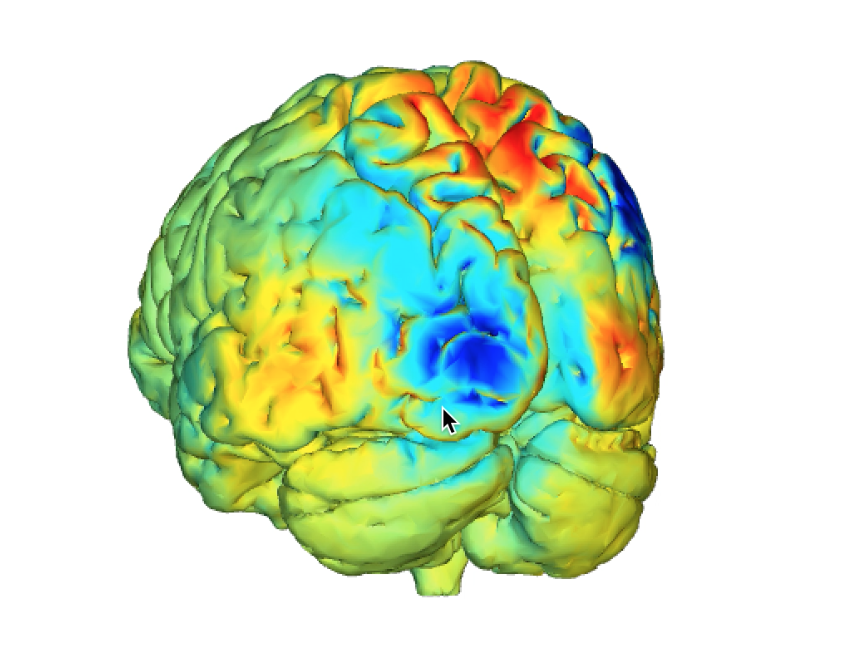}

%                \includegraphics[width=11cm,angle=0]{tomosample2}
%                
%\hspace{0.5cm}            \includegraphics[width=10cm,angle=0]{tomoresampled}
%            
    \caption{Sample tomographic solutions with 26 and 38 electrodes.  On the top, source dipole patch and reconstruction. On the bottom, reconstructed EEG from a sample subject (20 electrodes). Red colors represent outgoing sources, blue ones ingoing sources. } 
    
  \label{fig:reciprocity}

\end{figure}

%%%%%%%%%%%%%%%%%%%%%%%%%%%
%%%%%%%%%%%%%%%%%%%%%%%%%%%%%%%%%%%%%%%%%
\clearpage
\section{Application of reciprocity for  EEG-guided multichannel current brain stimulation (MtCS)}

Here we discuss how we can derive scalp potentials from EEG to design optimal stimulation protocols.

Using our tomographic approach, the EEG data can be inverted to source space to define a target map, and then optimize  as discussed [Ruffini2014\footnote{Ruffini et al., NeuroImage 89 (2014) 216Ð225}]. That is, from EEG we derive a cortical (or volume) dipole field map, and then we can specify that the generated electric fields by MtCS align or anti-align with them with specified strengths and so on. The advantage of this approach is that it allows for flexible specification of targets and of weight maps. Spontaneous EEG or ERP data can be pre-filtered to bands of interest or treated in other ways (ICA, etc).

Here we discuss a simpler approach (although a less flexible one as well).

Consider the theorem for two dipole sources measured again at a single scalp point $a$:
 \begin{equation}
 V_{a_1} \, I_{a} +V_{a_2} \, I_{a}   = - \vec{J}(x_1) \cdot  \vec{E}(x_2) \delta V + - \vec{J}(x_2) \cdot  \vec{E}(x_2) \delta V
\end{equation}
but this is just the resulting potential, of course, $V_a\, I_a$. Hence, it is clear that for arbitrary sources, we can generalize the reciprocity theorem a bit,
 \begin{equation}
 V_{a} \, I_{a}   = - \int dx \, \vec{J}(x) \cdot  \vec{E}(x) 
\end{equation}
We can see where this is going ... but let us get there faster.

Using the reciprocity relations we can use the lead matrix to relate a dipole source distribution in the brain to scalp potentials, 
$$
V_a=K_{ax}\, J(x)
$$
with implicit summation over repeated indices (vector indices on densities dropped for simplicity, also summed). The reference point for currents and potentials $b$ has been dropped for notational simplicity.

Here we recall that  $K=-\hat{E}_{ax}$, is the lead field matrix associated to unitary currents, mapping interior to scalp space (thanks to reciprocity).
Hence (summing over $x$ implicit), 
$$
V_a=-E_{ax}\, J(x)
$$
Let's multiply (and implicitly sum over repeated indices using the Einstein convention) both sides by a vector of injected currents $I_a$ (meaning $I_{ab}$):
$$
I_a\, V_a=-I_a \, E_{ax}\, J(x)
$$

Now, if we inject currents $I_{a}$, the generated electric fields are
$$
E(x)=\hat{E}_{ax} \, I_a
$$
Hence, 
$$
I_a\, V_a=-\, E(x)\cdot J(x)
$$
or, with the explicit summations written out
\begin{equation}
\sum_a I_a\, V_a = - \int E(x)\cdot J(x) dx
\end{equation}

Finally, we can imagine a future current stimulation system which is capable of controlling scalp current density. The corresponding equation would be 
\begin{equation}
\int_A  V_a \, dA\cdot J_a = - \int_V E(x)\cdot J(x) dx
\end{equation}

This rather elegant expression says that if you want generated electric fields and EEG sources to be correlated, currents and potentials have to be anti-correlated. This gives a simple way to determine optimal stimulation currents given scalp potential. Make currents and potentials to be as ``parallel'' as possible, given the constraints. E.g., maximize $|I_aV_a|$ subject to some constraints on maximal current, current sum, total injected current, etc.

We note that this will not optimize things like the electric field magnitude at the sources. Rather the specific component of the electric field parallel to the sources.

A limitation of this model-free optimization is that we will not really know what size electric fields we are generating, cannot add a weight map to work with weighted correlation, etc. But still, give some constraints on currents it provides a recipe to optimize currents to EEG sources. This can be especially useful in close-loop applications.

\subsection{Applications}
The reciprocity theorem above can be used for various applications, including
\begin{itemize}
\item Online optimization of MtCS from EEG (model-blind optimization based on EEG data)
\item Closed-loop applications from EEG: monitor EEG, create stimulation waveform (e.g., of the form I=c V or other optimizations as discussed above) to amplify or reduce EEG
\item Experiments involving ``Playing backÓ EEG using MtCS currents (I ? V), which may actually make some sense based on the above equation. 
\item Theoretical analysis purposes (e.g., EEG/tCS electrode density needed, see next section)
\end{itemize}

\subsection{Spatial resolution of tCS}
We can also use this equation to study the issue of how many electrodes (density) we need for stimulation. If we assume that the sources lie in a 2D surface (cortical mapping), for example, it is easy to argue that the map from sources to EEG is 1-1, and that, vice-versa, the map from scalp currents to cortical space E-fields is 1-1. Thus, in principle, if we could measure the full 2D surface with EEG (continuum EEG) we could invert the problem for sources, or if we could stimulate over all the entire scalp surface (continuum stimulation), we could reproduce any cortical target. Then, the needed number of EEG or stimulation electrodes would be transferred to the question of what the spatial scale of cortical sources or target is.  

In other words, the density of EEG or tCS electrodes needed for full inversion or target in this simplified model depends on the spatial scale (autocorrelation length) of the dipole sources on the cortex. We are ignoring here the issue of the finite size of electrodes, which act as a spatial filter. That is, we assume infinitely small electrodes. 

%%%%%%%%%%%%%%%
%\section{Matlab code}
% 
%
%
%\subsection{Matlab batch file: testesr.m}
%This is the  batch file to read E field files, compute the lead matrix and then inverse.
%{\footnotesize
%\lstinputlisting{CorticalMapDemo.m}
%}
%%%

%%%%%%%%%%%%%%%%%%%%%%%%%%%%%%%%%
%%%%%%%%%%%%%%%%%%%%%%%%%%%%%%%%%
%%%%%%%%%%%%%%%%%%%%%%%%%%%%%%%%%
%%%%%%%%%%%%%%%%%%%%%%%%%%%%%%%%%
%%%%%%%%%%%%%%%%%%%%%%%%%%%%%%%%%
%%%%%%%%%%%%%%%%%%%%%%%%%%%%%%%%%
%%%%%%%%%%%%%%%%%%%%%%%%%%%%%%%%%
%%%%%%%%%%%%%%%%%%%%%%%%%%%%%%%%%
%%%%%%%%%%%%%%%%%%%%%%%%%%%%%%%%%
%%%%%%%%%%%%%%%%%%%%%%%%%%%%%%%%%
%%%%%%%%%%%%%%%%%%%%%%%%%%%%%%%%%
%%%%%%%%%%%%%%%%%%%%%%%%%%%%%%%%%
%%%%%%%%%%%%%%%%%%%%%%%%%%%%%%%%

%\bibliographystyle{abbrv}  
%\bibliography{/Users/giulio/BIB/kolmogorov.bib}  % sigproc.bib is the name of the Bibliography in this case
% You must have
\end{document}